\def\gapprox{{_>\atop{^\sim}}}
\def\lapprox{{_<\atop{^\sim}}}

\def\cmsq{\rm {cm^2}}
\def\s-1{\rm {s^{-1}}}
\def\twco{$^{12}$CO}
\def\thco{$^{13}$CO}

\def\jb{\,Jy\,beam$^{-1}$}
\def\mjb{\,mJy\,beam$^{-1}$}
\def\rco{${\cal R}_{12/13}$}
\documentclass{aa}
\usepackage{graphics}
\begin{document}
\thesaurus{03(11.09.1, 11.09.4, 11.19.3, 13.19.1) }
%
%
%
\def\etal {et al.}
\def\ie {i.\,e.}
\def\etseq {{\em et seq.}}
\def\vs {{it vs.}}
\def\perse {{it per se}}
\def\adhoc {{\em ad hoc}}
\def\eg {e.\,g.}
\def\etc {etc.}
\def\ccpers {\hbox{${\rm cm}^3{\rm s}^{-1}$}}
\def\DEGR {\hbox{$^{\circ }$}}
\def\vlsr {\hbox{${v_{\rm LSR}}$}}
\def\vel {\hbox{${v_{\rm LSR}}$}}
\def\vhel {\hbox{${v_{\rm HEL}}$}}
\def\delv {\hbox{$\Delta v_{1/2}$}}
\def\dvel {\hbox{$\Delta v_{1/2}$}}
\def\TL {$T_{\rm L}$}
\def\TC {$T_{\rm c}$}
\def\TEX {$T_{\rm ex}$}
\def\TMB {$T_{\rm MB}$}
\def\TKIN {$T_{\rm kin}$}
\def\TREC {$T_{\rm rec}$}
\def\TSYS {$T_{\rm sys}$}
\def\TVIB {$T_{\rm vib}$}
\def\TROT {$T_{\rm rot}$}
\def\TDUST {$T_{\rm d}$}
\def\TASTAR {$T_{\rm A}^{*}$}
\def\TVIBST {$T_{\rm vib}^*$} 
\def\TB {$T_{\rm B}$}
\def \la{\mathrel{\mathchoice   {\vcenter{\offinterlineskip\halign{\hfil
$\displaystyle##$\hfil\cr<\cr\sim\cr}}}
{\vcenter{\offinterlineskip\halign{\hfil$\textstyle##$\hfil\cr
<\cr\sim\cr}}}
{\vcenter{\offinterlineskip\halign{\hfil$\scriptstyle##$\hfil\cr
<\cr\sim\cr}}}
{\vcenter{\offinterlineskip\halign{\hfil$\scriptscriptstyle##$\hfil\cr
<\cr\sim\cr}}}}}
\def \ga{\mathrel{\mathchoice   {\vcenter{\offinterlineskip\halign{\hfil
$\displaystyle##$\hfil\cr>\cr\sim\cr}}}
{\vcenter{\offinterlineskip\halign{\hfil$\textstyle##$\hfil\cr
>\cr\sim\cr}}}
{\vcenter{\offinterlineskip\halign{\hfil$\scriptstyle##$\hfil\cr
>\cr\sim\cr}}}
{\vcenter{\offinterlineskip\halign{\hfil$\scriptscriptstyle##$\hfil\cr
>\cr\sim\cr}}}}}
\def\RZWCO {${\cal R}_{2/1}^{\rm C^{18}O}$}
\def\RDRCO {${\cal R}_{3/2}^{\rm C^{18}O}$}
\def\RMSIO {${\cal R}_{5/2}^{\rm ^{28}SiO}$}
\def\RISOSIO {${ r }_{28/29}^{\rm SiO}$}
\def\RISOZSIO {${ r}_{29/30}^{\rm SiO}$}
\def\H0 {$H_{\rm o}$}
\def\mic {$\mu\hbox{m}$}
\def\micro {\mu\hbox{m}}
\def\SDOZ {\hbox{$S_{12\mu \rm m}$}}
\def\STWE {\hbox{$S_{25\mu \rm m}$}}
\def\SSIX {\hbox{$S_{60\mu \rm m}$}}
\def\SHUN {\hbox{$S_{100\mu \rm m}$}}
\def\solmass {\hbox{M$_{\odot}$}}
\def\solum {\hbox{L$_{\odot}$}}
\def\irlum {\hbox{$L_{\rm IR}$}}
\def\ohlum {\hbox{$L_{\rm OH}$}}
\def\blum {\hbox{$L_{\rm B}$}}
\def\numd {\hbox{$n\,({\rm H}_2$)}}                   
\def\rhounit {$\hbox{M}_\odot\,\hbox{pc}^{-3}$}
\def\kms {\hbox{${\rm km\,s}^{-1}$}}
\def\kmsyr {\hbox{${\rm km\,s}^{-1}\,{\rm yr}^{-1}$}}
\def\kmsmpc {\hbox{${\rm km\,s}^{-1}\,{\rm Mpc}^{-1}$}} 
\def\Kkms {\hbox{${\rm K\,km\,s}^{-1}$}}
\def\percc {$\hbox{{\rm cm}}^{-3}$}    
\def\cmsq  {$\hbox{{\rm cm}}^{-2}$}    
\def\cmsix  {$\hbox{{\rm cm}}^{-6}$}  
\def\arcsec {\hbox{$^{\prime\prime}$}}
\def\arcmin {\hbox{$^{\prime}$}}
\def\ffam {\hbox{$\,.\!\!^{\prime}$}}
\def\ffas {\hbox{$\,.\!\!^{\prime\prime}$}}
\def\ffM {\hbox{$\,.\!\!\!^{\rm M}$}}
\def\ffm {\hbox{$\,.\!\!\!^{\rm m}$}}
\def\ffs {\hbox{$\,.\!\!^{\rm s}$}}
\def\ffd {\hbox{$\,.\!\!^{\circ}$}}
\def\HI  {\hbox{HI}}
\def\HII {\hbox{HII}}
%
%
\def \AL {$\alpha $}    
\def \BE {$\beta $}     
\def \GA {$\gamma $}    
\def \DE {$\delta $}    
\def \EP {$\epsilon $}  
\def \alde {($\Delta \alpha ,\Delta \delta $)}
\def \MU {$\mu $}       
\def \TAU {$\tau $}     
\def \tapp {$\tau _{\rm app}$}
\def \tuns {$\tau _{\rm uns}$}
\def \RH {\hbox{$R_{\rm H}$}}         
\def \RT {\hbox{$R_{\rm \tau}$}}      
\def \BN  {\hbox{$b_{\rm n}$}}        
\def \BETAN {\hbox{$\beta _n$}}       
\def \TE {\hbox{$T_{\rm e}$}}         
\def \NE {\hbox{$N_{\rm e}$}}         
%
\def\MOLH {\hbox{${\rm H}_2$}}                    
\def\HDO {\hbox{${\rm HDO}$}}                     
\def\AMM {\hbox{${\rm NH}_{3}$}}                  
\def\NHTWD {\hbox{${\rm NH}_2{\rm D}$}}           
\def\CTWH {\hbox{${\rm C_{2}H}$}}                 
\def\TCO {\hbox{${\rm ^{12}CO}$}}                 
\def\CEIO {\hbox{${\rm C}^{18}{\rm O}$}}          
\def\CSEO {\hbox{${\rm C}^{17}{\rm O}$}}          
\def\CTHFOS {\hbox{${\rm C}^{34}{\rm S}$}}        
\def\THCO {\hbox{$^{13}{\rm CO}$}}                
\def\WAT {\hbox{${\rm H}_2{\rm O}$}}              
\def\WATEI {\hbox{${\rm H}_2^{18}{\rm O}$}}       
\def\CYAN {\hbox{${\rm HC}_3{\rm N}$}}            
\def\CYACFI {\hbox{${\rm HC}_5{\rm N}$}}          
\def\CYACSE {\hbox{${\rm HC}_7{\rm N}$}}          
\def\CYACNI {\hbox{${\rm HC}_9{\rm N}$}}          
\def\METH {\hbox{${\rm CH}_3{\rm OH}$}}           
\def\MECN {\hbox{${\rm CH}_3{\rm CN}$}}           
\def\METAC {\hbox{${\rm CH}_3{\rm C}_2{\rm H}$}}  
\def\CH3C2H {\hbox{${\rm CH}_3{\rm C}_2{\rm H}$}} 
\def\FORM {\hbox{${\rm H}_2{\rm CO}$}}            
\def\MEFORM {\hbox{${\rm HCOOCH}_3$}}             
\def\THFO {\hbox{${\rm H}_2{\rm CS}$}}            
\def\ETHAL {\hbox{${\rm C}_2{\rm H}_5{\rm OH}$}}  
\def\CHTHOD {\hbox{${\rm CH}_3{\rm OD}$}}         
\def\CHTDOH {\hbox{${\rm CH}_2{\rm DOH}$}}        
\def\CYCP {\hbox{${\rm C}_3{\rm H}_2$}}           
\def\CTHHD {\hbox{${\rm C}_3{\rm HD}$}}           
\def\HTCN {\hbox{${\rm H^{13}CN}$}}               
\def\HNTC {\hbox{${\rm HN^{13}C}$}}               
\def\HCOP {\hbox{${\rm HCO}^+$}}                  
\def\HTCOP {\hbox{${\rm H^{13}CO}^{+}$}}          
\def\NNHP {\hbox{${\rm N}_2{\rm H}^+$}}           
\def\CHTHP {\hbox{${\rm CH}_3^+$}}                
\def\CHP {\hbox{${\rm CH}^{+}$}}                  
\def\ETHCN {\hbox{${\rm C}_2{\rm H}_5{\rm CN}$}}  
\def\DCOP {\hbox{${\rm DCO}^+$}}                  
\def\HTHP {\hbox{${\rm H}_{3}^{+}$}}              
\def\HTWDP {\hbox{${\rm H}_{2}{\rm D}^{+}$}}      
\def\CHTWDP {\hbox{${\rm CH}_{2}{\rm D}^{+}$}}    
\def\CNCHPL {\hbox{${\rm CNCH}^{+}$}}             
\def\CNCNPL {\hbox{${\rm CNCN}^{+}$}}             
%
%
\def\In {\hbox{$I^{n}(x_{\rm k},y_{\rm k},u_{\rm l}$})}
\def\Iobs {\hbox{$I_{\rm obs}(x_{\rm k},y_{\rm k},u_{\rm l})$}}
\def\Ingl {I^{n}(x_{\rm k},y_{\rm k},u_{\rm l})}
\def\Iobsgl {I_{\rm obs}(x_{\rm k},y_{\rm k},u_{\rm l})}
\def\Pbgl {P_{\rm b}(x_{\rm k},y_{\rm k}|\zeta _{\rm i},\eta _{\rm j})}
\def\Pbgm {P(x_{\rm k},y_{\rm k}|r_{\rm i},u_{\rm l})}
\def\Pbgn {P(x,y|r,u)}
\def\Pugm {P_{\rm u}(u_{\rm l}|w_{\rm ij})}
\def\Pdem {P_{\rm b}(x,y|\zeta (r,\theta ),\eta (r,\theta ))} 
\def\Pden {P_{\rm u}(u,w(r,\theta ))}
\def\greekgl {(\zeta _{\rm i},\eta _{\rm j},u_{\rm l})}
\def\greekg1 {(\zeta _{\rm i},\eta _{\rm j})}
\title{Changing Molecular Gas Properties in the Bar and Center of NGC~7479}
%
\author{S.~H\"{u}ttemeister\inst{1}, S.~Aalto\inst{2}, M.~Das\inst{3,4}
W.\ F.\ Wall\inst{5} }
\offprints{S.\ H\"uttemeister (huette@astro.uni-bonn.de)}
\institute{
 Radioastronomisches Institut der Universit\"{a}t Bonn,
 Auf dem H\"{u}gel 71, D - 53121 Bonn, Germany
\and
 Onsala Space Observatory, S - 43992 Onsala, Sweden
\and
Indian Institute of Astrophysics, Bangalore, India
\and
Department of Astronomy, University of Maryland, College Park, MD 20742, USA 
\and
 INAOE, 72000 Puebla, Mexico
 }
\date{Received 19 March 2000 / Accepted 24 August 2000 }
\titlerunning{Gas properties in the bar of NGC\,7479}
\authorrunning{H\"uttemeister et al.}
\maketitle
\begin{abstract}
We present sensitive interferometric $^{12}$CO, $^{13}$CO and HCN 
observations of the barred spiral galaxy NGC\,7479, known to be one of the
few barred galaxies with a continuous gas-filled bar. We focus on the 
investigation and interpretation of $^{12}$CO/$^{13}$CO line intensity
ratios \rco , which is facilitated by having more than 90\%
of the flux in 
our interferometer maps. The global (9\,kpc by 2.5\,kpc) value of \rco\ 
is high at 20 -- 40.  On smaller scales ($\sim 750$\,pc), \rco\ is found 
to vary dramatically, reaching values $> 30$ in large parts of the bar, 
but dropping to values $\sim 5$, typical for galactic disks, at a 
$^{13}$CO condensation in the southern part of the bar. 
We interpret these changes in terms of the 
relative importance of the contribution of a diffuse molecular component,
characterized by unbound gas that has a moderate optical depth in the 
$^{12}$CO(1$\to$0) transition. This component dominates the $^{12}$CO
along the bar and is also likely to play an important role in the 
center of NGC\,7479. In the center, the $^{12}$CO and the HCN intensity
peaks coincide, while the $^{13}$CO peak is slightly offset. This can
be explained in terms of high gas temperature at the $^{12}$CO peak 
position. 
Along the bar, the relation between the distribution of $^{12}$CO, $^{13}$CO,
dust lanes and velocity gradient is complex. A southern $^{13}$CO condensation
is found offset from the $^{12}$CO ridge that generally coincides 
with the most prominent dust lanes. It is possible that strong $^{13}$CO
detections along the bar indicate quiescent conditions, downstream from 
the major bar shock. Still, these condensations are found close to 
high velocity gradients. In the central region, the velocity gradient
is traced much more closely by $^{13}$CO than by $^{12}$CO.  
\end{abstract} 
\section{Introduction}
Bars are regarded as an important transport mechanism of material toward
the central regions of galaxies, fuelling nuclear starbursts and 
driving galaxy evolution through the concentration of mass close to the
nucleus (e.g.\ Sakamoto et al.\ \cite{saka+}, Combes \cite{combes+} 
and references therein). To accomplish this, gas has to flow inward through 
the bar; however, long bars that are filled with dense gas that can 
be traced in CO are rare
and thus likely transient. Among the few specimens known are NGC\,1530
(Reynaud \& Downes \cite{rendow}), M\,100 (Sempere \& Garc\'{\i}a-Burillo 
\cite{sem_ga}), UGC\,2855 (H\"uttemeister et al.\ \cite{huette+}),
NGC\,2903 and NGC\,3627 (Regan et al.\ \cite{regan+}). These bars 
differ significantly in their properties, i.e.\ 
velocity field, linewidth and derived shock structure. This is somewhat 
surprising, since the gas should in all cases be responding to a strong 
bar potential in a similar way. The dynamics of this response should 
then have consequences for the properties of the gas in the bar and 
the nucleus, e.g.\ produce a diffuse component unbound from clouds
(e.g.\ Das \& Jog \cite{dasjog}). The evolutionary state of the bar and 
the degree of central mass concentration may be instrumental in regulating 
the conditions of the gas in the bar (see models by Athanassoula 
\cite{atha}).  More specifically, the gas is funnelled from outer 
$x_1$-orbits to inner $x_2$-orbits as the bar evolves in time (see e.g.\
the simulations by Friedli \& Benz \cite{frie_be}).

\begin{figure*}
\resizebox{21cm}{!}{\includegraphics{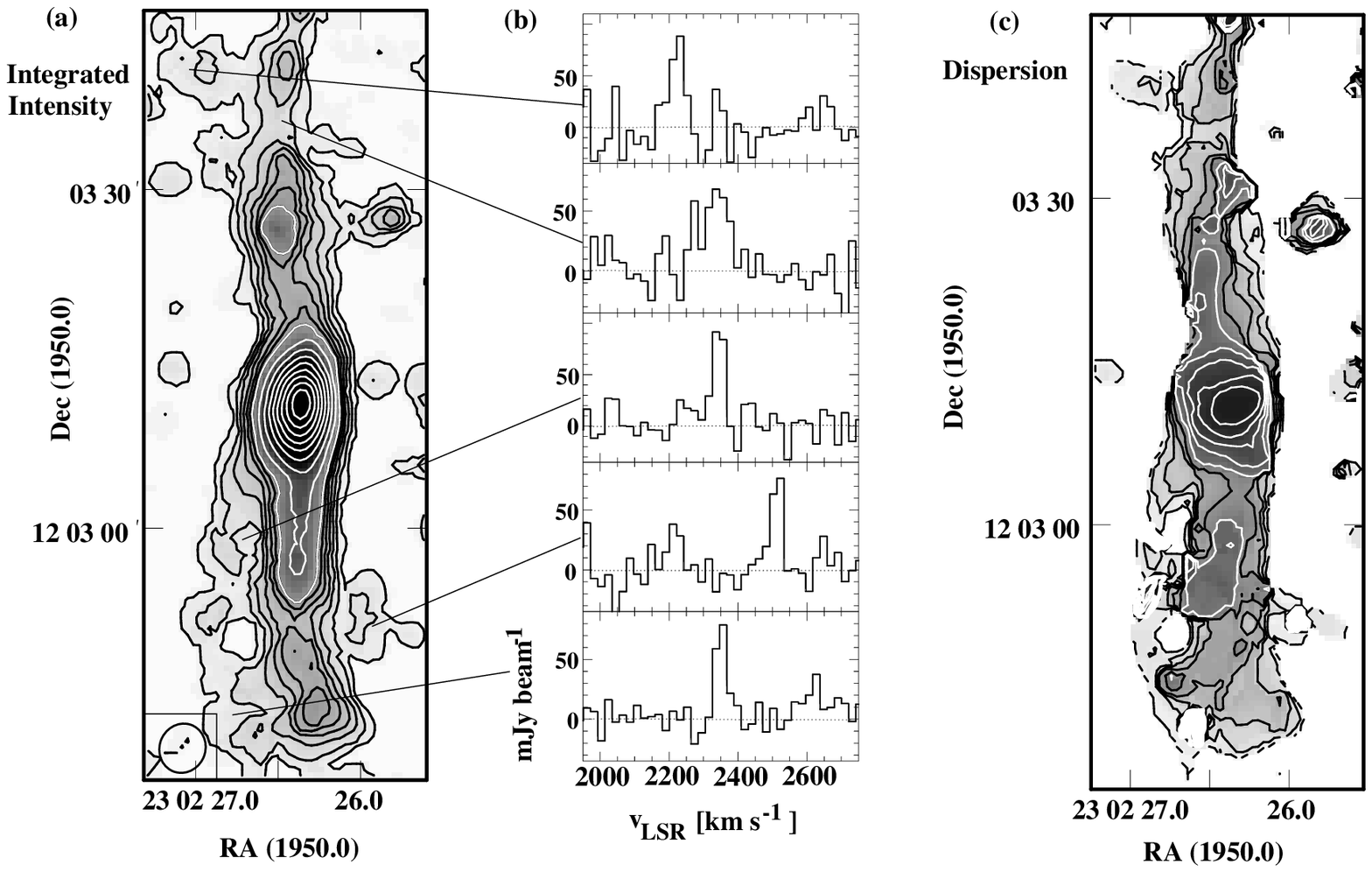}}
\caption{\label{12co} (a) The map of the $^{12}$CO (1 $\to$ 0) total 
integrated intensity. The contour levels range from 7.6\,K\,\kms\ 
(1.6\,\jb\,\kms\ or 1.5$\sigma$ for a velocity width of 400\,\kms ) to   
76\,K\,\kms\ (16\,\jb\,\kms ) in steps of 11.4\,K\,\kms\ (2.4\,\jb\,\kms ), 
2$\sigma$) (black), then change to a spacing of 38\,K\,\kms\  
(8.0\,\jb\,\kms\ , 8$\sigma$) up to 380\,K\,\kms\ (80\,\jb\,\kms ) (white). 
(b) Representative spectra from the outer regions of the maps, 
demonstrating that the lines in these regions become narrow, but are 
clearly detected. The spectra correspond to single beams centered on the
positions indicated. (c) Dispersion map. The contours range from 10\,\kms\ 
to 100\,\kms\ in steps of 10\,\kms . The peak dispersion is 110\,\kms .}
\end{figure*}

To better understand the gas flow, the star formation efficiency along 
the bar and the feeding of a starburst, it is essential to know the  
physical state of the gas both along the bar and in the central 
structure. An analysis based on more molecules than $^{12}$CO, the most
abundant species after H$_2$, is necessary to investigate gas properties 
and to go beyond the study of the
morphological and kinematic structure of a system. Achieving this goal
requires the detection and high-resolution mapping of very 
faint lines, observations which are at the sensitivity limit of today's 
millimeter interferometers. 

In this paper, we present such a study, carried out in one of the best 
known gas-rich barred galaxies. NGC\,7479, classified a (mild) starburst 
SBb galaxy containing a LINER nucleus, has a gaseous bar of $\sim
10$\,kpc projected length (for an assumed distance of $D=32$\,Mpc, 
see Table\,\ref{prop}), which
has been repeatedly mapped in $^{12}$CO: Sempere et al.\ 
(\cite{sempere+}) present single dish data in the $1 \to 0$ and $2 \to 1$
transition. Interferometric $1 \to 0$ maps have been obtained by 
Quillen et al.\ (\cite{quillen+}) and recently by Laine et al.\ 
(\cite{laine+}). Our results include the most sensitive $^{12}$CO map to 
date, but we focus on the relation of the $^{12}$CO emission to the first
interferometric $^{13}$CO$(1 \to 0)$ and HCN($1 \to 0$) data obtained 
in NGC\,7479, and the physical state of the gas we can infer from the
comparison of these molecules to the $^{12}$CO distribution. 
Preliminary results of this study were given in Aalto et al.\ (\cite{proc}).

\begin{table*}
\caption{\label{prop} Adopted properties of NGC\,7479}
\begin{tabular}{rrrrrrrrrr}
\multicolumn{1}{c}{R.A. (1950.0)$^{\rm a)}$} 
& \multicolumn{1}{c}{Dec (1950.0)$^{\rm a)}$} 
& Distance 
& Type$^{\rm b)}$ & $m_b^{\rm b)}$ 
& $i^{\rm c)}$ & Size (blue)$^{\rm b)}$
& $L_{\rm FIR}^{\rm d)}$ & PA disk$^{\rm e)}$ & PA bar$^{\rm e)}$ \\
\hline \\
 $23^{\rm h}\ 02^{\rm m}\ 26.37^{\rm s}$ & $12^{\circ}\ 03'\ 
10.6''$  & 32 Mpc 
& SBb & 11\ffm 7$^{\rm c)}$ & $\sim 45^{\circ}$ 
& 4\ffam 4 $\times$ 3\ffam 4 & $4 \cdot 10^{10}$\solum 
& 25$^{\circ}$ & 5$^{\circ}$ \\

\hline 
\end{tabular} \\
a): Radio continuum peak position (Neff \& Hutchings \cite{nefhut},
VLA A-Array at 1.490\,MHz) \\
b): from the UGC catalogue (Nilson \cite{nilson}) \\
c): from Groosbol (\cite{grosbol}) $i$: angle of inclination \\
d): from Sempere et al.\ (\cite{sempere+})\\
e): Position angle (north-to-east) of the disk major axis (see e.g.\
 the UGC catalogue) and 
the gaseous bar. The position angle of the stellar bar is even closer to
that of the disk major axis (Sempere et al.\ \cite{sempere+}). 
\end{table*}

\section{Observations}

\subsection{Interferometric observations at OVRO}

We have obtained interferometric maps of the $1 \to 0$ transitions of
$^{12}$CO, $^{13}$CO, HCN and in the inner 1$'$ (9\,kpc) of NGC\,7479 
using the Caltech 
six-element Owens Valley Radio Observatory (OVRO) millimeter array. 
All observations were centered on the radio continuum peak. The quasar
3C454.5 was observed every $\sim 15$\,min as a phase and amplitude calibrator.
3C345 was used as an additional passband calibrator, and flux calibration
was done relative to the planets Neptune and Uranus. Calibration was done
with the standard package developped for OVRO. 
Continuum emission with a peak flux of 3 -- 4\,\mjb\ has been subtracted 
from all maps.

The observations were carried out in the low resolution and equatorial
configurations of the OVRO array between February and June 1996.
The naturally weighted synthesized beam sizes are 4\ffas 6 $\times$ 
4\ffas 25 for $^{12}$CO,  4\ffas 9 $\times$ 4\ffas 05 for \thco\ and 7\ffas 2 
$\times$ 5\ffas 2 for HCN. 
At $D$=32\,Mpc, $1''$ equals 155\,pc. The {\em positional accuracy}
of structures in the maps is $\sim 0\ffas 5$ and thus considerably 
higher than the resolution. Due to the limited $uv$-coverage, the
maps are not sensitive to structures larger than $\sim 15''$ for CO and 
$\sim 21''$ for HCN. On-source integration times were 7.2\,hr for \twco ,
23.5\,hr for \thco\ and 11.5\,hr for HCN.  

The parameters of the observations are summarized in Table\,\ref{obs}. 

For a beamsize of $4\ffas 4$, i.e.\ \twco , a brightness temperature 
(\TB ) of 1\,K corresponds to 0.210\,Jy beam$^{-1}$ at a wavelength of 
2.6\,mm. The spectral resolution of our data is 4\,MHz or 10.04\,\kms, 
and the total velocity range covered by the autocorrelator is 1120\,\kms ,
centered on \vlsr = 2500\,\kms. For the final presentation, the data 
were smoothed to a velocity resolution of 20\,\kms\ ($^{12}$CO and 
$^{13}$CO) or 40\,\kms\ (HCN). 

Imaging was done using the NRAO AIPS package. For image display and analysis
of the data cubes, the ATNF Karma package was also used. 

\begin{table*}
\caption{\label{obs} Summary of the interferometric observations of NGC\,7479}
\begin{tabular}{rlllcllll}
\multicolumn{1}{c}{Line, $\nu$ (GHz)} 
& \multicolumn{1}{c}{Date} 
& Field Center & $uv$-Coverage &  Prim.\ Beam & Syn.\ Beam
& Conf.$^{a)}$ & $T_{\rm sys} $ & Noise$^{b)}$ \\
\hline \\
$^{12}$CO($1\to 0)$ & 2/96  & 
 $23^{\rm h}\ 02^{\rm m}\ 26.4^{\rm s}$ &
 6k$\lambda$ -- 46\,k$\lambda$ & 63$''$ & $4\ffas 6 \times 4\ffas 25$ & L,E 
 & $\sim 500$K & 85\,mK \\
115.271 
 & 4/96 & $12^{\circ}\ 03'\ 10.6''$ & (15m -- 119m) &  
 & PA $-9^{\circ}$ & & \\
$^{13}$CO$(1\to 0)$ & 2/96, 3/96  &  & 5.7k$\lambda$ -- 44\,k$\lambda$ 
 & 66$''$ &  $4\ffas 9 \times 4\ffas 05$ & L,E & $\sim 300$K & 23\,mK  \\
110.201
 &  5/96,  6/96 & & (15m -- 119m) & & PA $-45^{\circ}$ & & \\
HCN$(1\to 0)$ & 2/96  &  & 4.4k$\lambda$ -- 35\,k$\lambda$
 & 80$''$ & $7\ffas 2 \times 5\ffas 2$  & L,E  & $ \sim 300$K & 12\,mK \\
88.632
 & 5/96 && (15m -- 119m) &  & PA $-49^{\circ}$ & & \\
\hline 
\end{tabular} \\
$a)$: The array configurations we used at OVRO are labelled L (low
resolution) and E (equatorial) \\
$b)$: This is the rms noise per channel (20\,\kms\ for CO, 40\,\kms\ for
HCN). On a \mjb-scale, the values correspond to 18\,\mjb (\twco ),
and 5\,\mjb (\thco\ and HCN). 
\end{table*}

\subsection{Single dish observations at OSO}

Single dish spectra of the $1 \to 0$ transitions of \twco\ and \thco\
were taken at the Onsala Space Observatory 20\,m telescope in 1997. The 
central position of NGC\,7479 was observed with a beamsize of 33$''$
and system temperatures of 500\,K -- 600\,K (\twco ) and 200\,K -- 300\,K
(\thco ). The spectra were smoothed to a velocity resolution of 30\,\kms . 

\section{Results}

\subsection{The Distribution of the $^{12}$CO Emission}

\subsubsection{Morphology and Central Kinematics} 

The \twco\ distribution is displayed in Fig.\,\ref{12co}a. It extends 
in one continuous structure over the whole length of the bar, with a
number of secondary peaks along the bar axis. The total $^{12}$CO
flux in the map is 380\,Jy\,\kms . The distribution has a central 
concentration with a (deconvolved) FWHM size of 
$7\ffas 5 \times 2\ffas 6$  ($1200 \times 
400$\,pc at $D$=32\,Mpc). This structure holds about 30 \% of the total 
CO flux of the bar. The peak position is at $\alpha = 23^{\rm h}\ 
02^{\rm m}\ 26.37^{\rm s}$ and $\delta = 12^{\circ}\ 03'\ 11\ffas 6$,
coincident to within 1\ffas 0 with the radiocontinuum peak.

\begin{figure*}
\resizebox{\hsize}{!}{\includegraphics{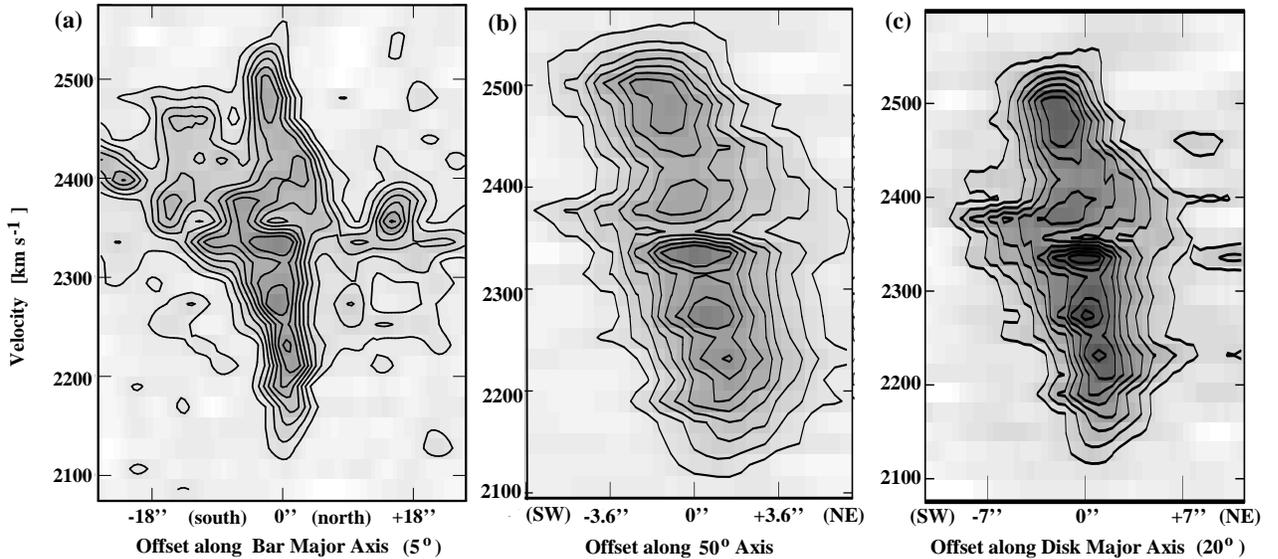}}
\caption{\label{pv} a) Position velocity diagram along the major axis of
the \twco\ bar (position angle 5$^{\circ}$ (north-to-east)). Thus, the
bar is oriented very close to the north-south direction. 
b) Position velocity diagram along the axis with 
position angle 50$^{\circ}$ (i.e.\ 45$^{\circ}$ relative to the bar major 
axis). This is the angle that best shows the central rapidly rotating 
structure. 
c) Position velocity diagram along the disk major axis, at a 
position angle of 25$^{\circ}$ (i.e.\ 20$^{\circ}$ relative to the bar 
major axis). For all panels, 
the contours are multiples of 2$\sigma$ (0.17\,K or 36\,\mjb ). }
\end{figure*}

In Fig.\,\ref{12co}b, we display some representative spectra from the
outer parts of our \twco\ map, regions where the confidence level of
the integrated intensity maps (which refers to the full velocity
width in the center of 400\,\kms ) drops below $2\sigma$. The lines are 
clearly detected in these parts of the bar, but become very narrow. The
same effect is evident from the dispersion map (Fig.\,\ref{12co}c):
The dispersion (calculated along the line-of-sight, $\delta v_{\rm mom}$) 
falls off sharply toward the bar edges, but remains at 40\,\kms\ to 60\,\kms\ 
close to the major axis. 

\begin{figure*}
\resizebox{\hsize}{!}{\includegraphics{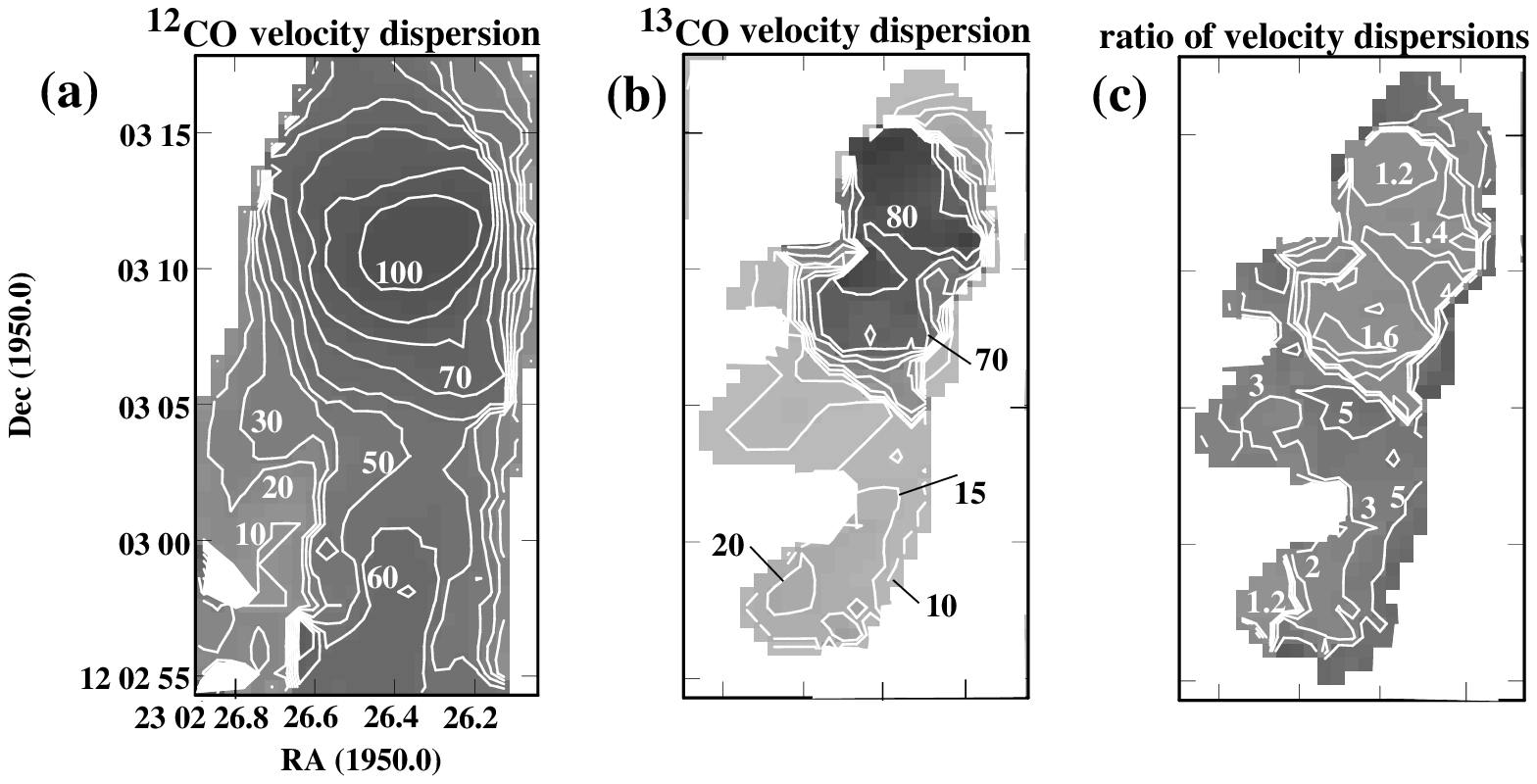}}
\caption{\label{disp} Maps of the one-dimensional velocity 
dispersion in the
inner region where continuous \thco\ emission is detected. a) \twco. 
The contours range from 10\,\kms\ to 100\,\kms\ in steps of 10\,\kms .
b) \thco. The contour levels are at 10\,\kms , 15\,\kms and then range
from 20\,\kms\ to 80\,\kms\ in steps of 10\,\kms . c) Ratio of the \twco\
velocity dispersion to the \thco\ velocity dispersion. The contours label
ratios of 1.2, 1.4, 1.6, 2, 3 and 5.}   
\end{figure*}

To show that the change in dispersion is not a result of projected
rotation, we display position-velocity (pv) diagrams along the bar major
axis (position angle (PA) 5$^{\circ}$, Fig.\,\ref{pv}a) and the disk major
axis (PA 25$^{\circ}$, Fig.\,\ref{pv}c). The velocity width is not 
systematically larger along the disk major axis, where rotation should have 
the largest effect. Also, the regions of narrow lines are not associated 
with the disk minor axis. The only possible effect of rotation is a tentative 
alignment of the 60\,\kms\ to 80\,\kms\ velocity contours in the 
\twco\ dispersion map (Fig.\,1c and Fig.\,\ref{disp}a) with the disk major 
axis. 

\begin{figure*}
\resizebox{\hsize}{!}{\includegraphics{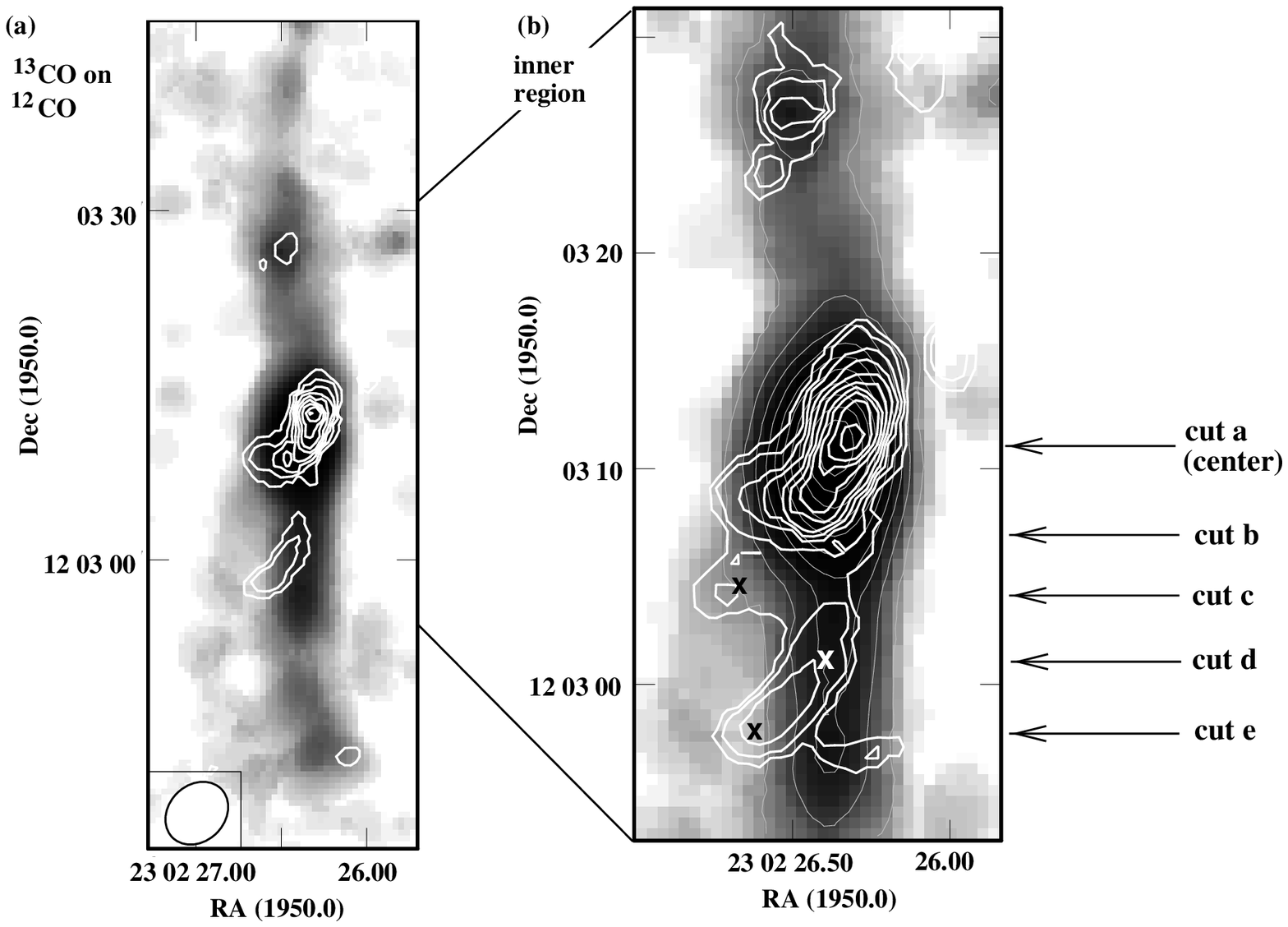}}
\caption{\label{13co} The distribution of the \thco\ emission (contours)
superposed on the \twco\ intensity map (greyscale). a) Intensity integrated
over the full velocity width in the center, i.e. 400\,\kms . The contours 
start at 1.4\,K\,\kms\ (0.3\,\jb\,\kms\ or $1.5\sigma$ for a width of 
80\,\kms ) and are spaced by 1.85\,K\,\kms . b) The inner region of the 
bar, integrated over 200\,\kms\ only to bring out emission from narrow lines. 
The contours start at 0.9\,K\,\kms\ (0.2\,\jb\,\kms, 1.5$\sigma$ for 
a velocity
width of 40\,\kms ) with a spacing of 1.35\,K\,\kms . The crosses mark the
positions for which spectra are shown in Fig.\,5. Thin gray 
contours refer to \twco . The arrows mark the positions of the intensity 
cuts in Fig.\,\ref{intcut}.}  
\end{figure*}

There is no sign of a `twin peaks' structure (as coined by Kenney et al.\  
\cite{ken+}) perpendicular to the bar, perhaps suggesting that there 
are no $x_2$ antibar orbits, occuring within an Inner Lindblad resonance 
(ILR). However, the question of the existence of an ILR in NGC\,7479 
is controversial and remains undecided: Quillen et al.\ (\cite{quillen+})
derived the existence of an ILR close to the nucleus, while in the models 
of Sempere et al.\ (\cite{sempere+}) no ILR exists. Laine et al.\ 
(\cite{lai+2}) argue that the bar perturbation close to the center is so 
strong that the issue cannot be resolved without non-linear orbit analysis. 
In any case, the ILR would be located within 700\,pc -- 800\,pc (4$''$ --
5$''$) from the nucleus; thus, we would barely resolve a `twin peak'
structure. 

Another possible signature of an ILR is a ring of material 
close to its positions, due to the expected pile-up of gas in the collision
region where $x_1$ and $x_2$ orbits intersect.  Such a structure may be
indicated in our map. There certainly is a clear peak not just in intensity
but also in velocity dispersion (Fig.\,\ref{12co}c) in the center. This
dispersion peak fragments into a high velocity system, consisting of
a number of clumps, as is evident from the pv diagrams we 
show in Fig.\,\ref{pv}. This system is confined to within $\pm 3''$ of the
center and distinctly different from the material within the `main' bar,
where the velocity changes much more slowly, which is seen in panel\,(a),
a cut along the major bar axis. The central rapidly rotating structure is most
easily seen at a position angle close to $45^{\circ}$ (panel (b)). 
While there is 
a central peak (at $v_{\rm LSR} = 2330$\,\kms ), this peak does not
dominate the emission: other features at velocities ranging from 
2200\,\kms\ to 2500\,\kms\ are almost as strong and spatially separated by
$1'' - 2''$. 

A high velocity system like this is a signature feature of many 
barred galaxies. It has been used by Binney et al.\ (\cite{binney+}) to
argue the presence of a central bar in our Galaxy. The orbits of the gas 
that can cause this feature have been modelled by e.g.\ Garc\'{\i}a-Burillo
\& Guelin (\cite{gb}) for the case of the weakly barred edge-on
galaxy NGC\,891. 
The structure found in NGC\,7479 (already noted by Laine et al.\ 
\cite{laine+}) also greatly resembles the system seen in UGC\,2855, 
a strongly barred galaxy where, in contrast to NGC\,7479, quiescent 
conditions are found for the gas along the bar (H\"uttemeister et al.\
\cite{huette+}). As in UGC\,2855, the central high velocity system in 
NGC\,7479 can be interpreted as a clumpy tilted ring or torus close to the 
ILR (which we would then place at $2'' - 3''$ from the nucleus),  as 
a fragmented, tilted, rotating nuclear disk that is freely fed 
from material inflowing through the bar without being stopped at an ILR,
or even as a dynamically decoupled inner bar, as tentatively suggested
by Baker (\cite{baker}).

\subsubsection{Comparison with Prior $^{12}$CO Maps}

The general morphological structure and velocity field of our map are in 
excellent agreement with the results reported by 
Laine et al.\ (\cite{laine+}).
However, the $^{12}$CO emission in our map is a continuous bar structure,
while the structure seen by Laine et al.\ breaks up into separate clumps 
away from the central condensation, a large part of the body of the bar 
being devoid of emission. As can be expected if the two maps are
consistent, the position of these clumps is in exact agreement
with the peaks we see embedded in more extended emission. 

Comparing the total fluxes contained in the maps, we find that
our flux is higher by a factor of 1.45 than that seen by Laine et al. 
This discrepancy is caused by our map being more sensitive by a factor 
of $\sim 2$, due to a larger beam size combined with a slightly
longer integration time. Thus, the larger East-West diameter of the bar
seen in our map 
and its continuity are not artifacts of smearing by lower resolution, but 
real effects: Our map picks up an additional, more smoothly distributed gas 
component traced by $^{12}$CO. We can compare our total flux to the flux
seen in the single dish map of Sempere et al.\ (\cite{sempere+}), which
covered a larger area than our map. However, $\sim 80$\% of the flux 
is contained the region of the OVRO primary beam. We find that we recover 
more than 90\% of this flux, which is consistent with Laine et al., who
report recovering 60\% of the single dish flux in their less 
sensitive map. Thus, {\em there is very little, if any, missing flux 
in our \twco\ map} and we can analyse the complete molecular distribution
as traced by \twco\ at high resolution. {\it This complete recovery of
flux is the main distinguishing feature between our \twco\ map and 
the interferometric maps published previously}, both by Laine et al.\ 
and earlier by Quillen et al.\ (\cite{quillen+}). 

\begin{figure}
\resizebox{\hsize}{!}{\includegraphics{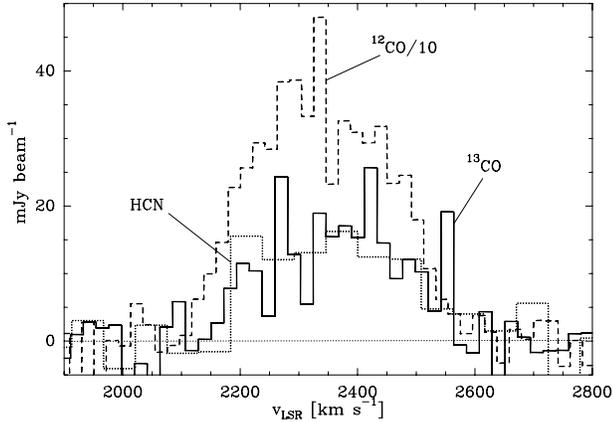}}
\caption{\label{spcen} Spectra of the central position (taken at the position
with the highest \twco\ intensity, i.e.\ 
$\alpha = 23^{\rm h}\ 02^{\rm m}\ 26.37^{\rm s}$ and 
$\delta = 12^{\circ}\ 03'\ 11\ffas 6$ ). 
The dashed line is \twco\ divided by 10 for better comparability, the 
solid line is \thco\ and the dotted line HCN. All three spectra above
correspond to a region with the 7\ffas 2 $\times$ 5\ffas 2 size of the 
HCN beam; the \twco\ and \thco\ data were smoothed to the same beamsize as
the HCN data. 20\,\mjb\ correspond to 73\,mK (HCN) or 44\,mK (\twco ).}
\end{figure}

\begin{figure}
\resizebox{\hsize}{!}{\includegraphics{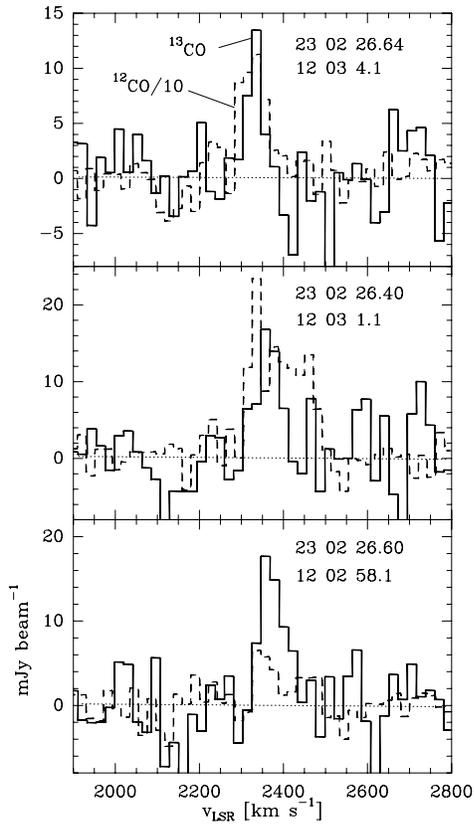}}
\caption{\label{spoff} Offset spectra for \twco\ divided by 10 (dashed)
and \thco\ (solid). The positions are marked in Fig\,\ref{13co} and 
correspond to regions with the size of the beam of the CO observations
($\sim 5'' \times 4''$, see Table\,2).
For the \twco\ beam used here, 20\,\mjb\ correspond to 97\,mK.}
\end{figure}

\subsection{Distribution of $^{13}$CO and HCN}

\begin{figure}
\resizebox{\hsize}{!}{\includegraphics{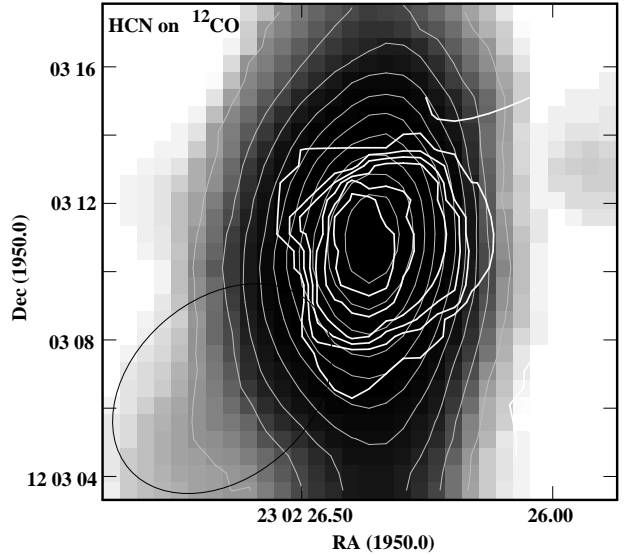}}
\caption{\label{hcn} The total integrated intensity of HCN, superposed
on \twco\ (grayscale and thin grey contours). The HCN contour levels start
at 1\,K\,\kms\ (0.42\,\jb\,\kms\ or 1.5$\sigma$) and increase in 2$\sigma$
steps.  The ellipse in the lower left corner denotes the HCN beam and 
demonstrates that the HCN emission is not resolved.} 
\end{figure}

Since \thco\ and HCN usually originate from regions that are more compact
and confined than \twco\ emitting gas, we consider it very likely that
we also recover the entire flux for those molecules, being limited only
by sensitivity, not by missing zero-spacings in the interferometer map. 

The distribution of the \thco\ emission is presented in Fig.\,\ref{13co}.
While Fig.\,\ref{13co}a gives the moment\,0 map for the full
central velocity of 400\,\kms\ (see Fig.\,\ref{spcen}), panel b is 
integrated over only 200\,\kms , so that very narrow lines are not lost 
in the noise. Now, a connection between the central distribution and 
the southern feature becomes apparent. The example spectra shown in 
Fig.\,\ref{spoff} demonstrate  the narrowness of these lines. 
The total \thco\ flux we detect is $\sim 9$\,Jy\,\kms .

The \thco\ emission is distributed very differently from the \twco .
The central \thco\ peak is significantly offset to the north-west from the 
\twco\ peak (by 2$''$ or 320\,pc in linear size).

\begin{figure}
\resizebox{\hsize}{!}{\includegraphics{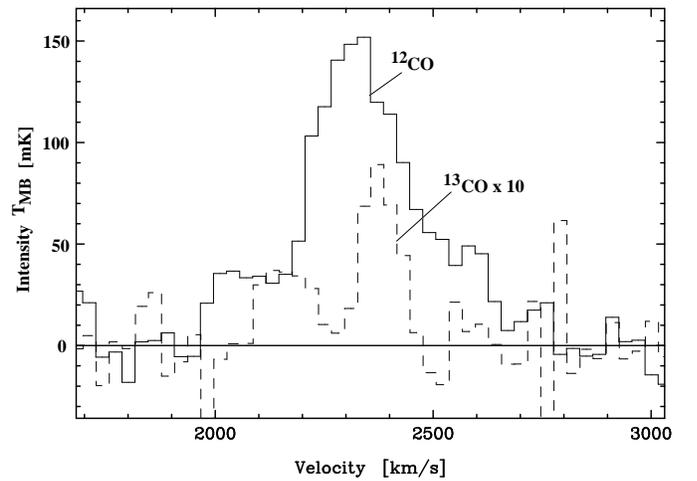}}
\caption{\label{sd} Single dish observations of \twco\ and \thco\ taken
with the OSO 20\,m telescope. The beamsize is 33$''$, the intensity scale
is $T_{\rm MB}$ (main beam efficiency $\eta_{\rm MB} = 0.5$).} 
\end{figure}

While the \twco\ distribution in the central concentration and south of 
the center along the bar runs almost exactly north-south, the central
\thco\ distribution is best described as extending along a position
angle of $- 30^{\circ}$, with a curvature toward the south-east 
(Fig\,\ref{13co}). The deviation between the \twco\ and \thco\
morphology is even clearer in the southern \thco\ extension at 
offsets $7''$ to $15''$ to the south along the bar. 

No \thco\ is seen at corresponding offset along in the northern part of
the the bar. 

The weak feature offset $\sim 20''$ to the north, close to the bar end,
has no southern counterpart, but is coincident with a peak in the
\twco\ distribution and thus likely to be real. 
Further peaks at the bar ends are not detected
with at a sufficient confidence level, since these regions are close to 
the edges of our primary beam, and any emission may be affected by 
the fall-off in sensitivity as well as sidelobes of the central peak. 
One might expect to recover \thco\ again close to the bar edges, due 
to the crowding of elliptical ($x_1$) streamlines, resulting in increased
cloud collisions and the formation of dense shocked gas. Further high 
sensitivity studies are needed to clarify this point. 

Still, very dramatic {\em changes in the \twco\ to \thco\ line
intensity ratio} are evident, both in the central region and along the
bar. These changes, which are indicators of changing gas properties, will
be discussed in detail in the subsequent section. 

HCN is clearly detected only in the very center of the \twco\ distribution
(Fig.\,\ref{hcn}). It is unresolved and its position is consistent 
with the \twco\ peak, i.e.\ offset from the maximum \thco\ intensity.
The total HCN flux in this central structure is 1.6\,Jy\,kms .

\subsection{Line intensity ratios}
\begin{figure}[h!]
\resizebox{12cm}{!}{\includegraphics{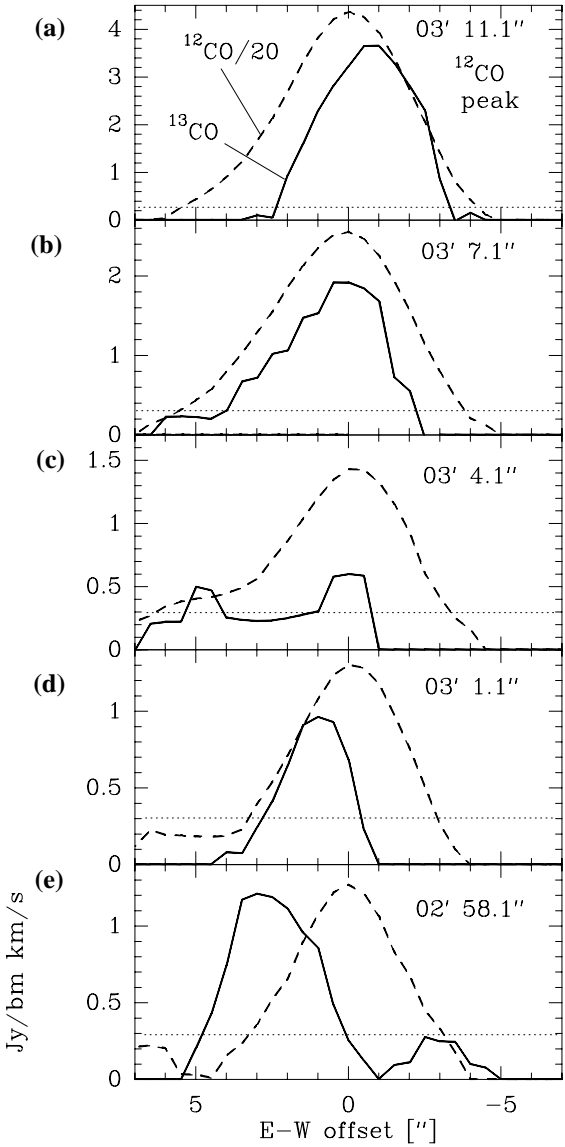}}
\caption{\label{intcut} East-West cuts in integrated intensity through the bar
at the positions marked in Fig.\,\ref{13co}. The north-south width of the
cuts correspond to the $4'' - 5''$ beamwidth of the CO observations.
Solid lines are \thco ,
dashed lines are \twco\ divided by 20. The integrated intensities refer
to a velocity range of 450\,\kms , i.e.\ the full range where \twco\ 
emission is found. The thin dashed line is the 2$\sigma$
limit for \thco\ and a velocity width (FWHM) of 40\,\kms , i.e.\ the
typical width of off-center \thco\ lines. Thus lower \thco\ 
intensities have no meaning. All offsets are relative to the right 
ascension of the \twco\ peak position. 1\,\jb\,\kms\ corresponds to 
4.8\,K\,\kms .  } 
\end{figure}

We present line intensity ratios taken between \twco\ and \thco\ 
(\rco ) in different positions and over different areas 
in Table\,\ref{ratios}. In this table, we have computed \rco\ and its
errors individually for the peak intensity, the integrated intensity as
measured within the velocity window where \thco\ is detected and the velocity
range within which we find \twco\ emission. The respective velocity ranges
are given along with the values of \rco . Again, it is obvious that the
\thco\ line usually is narrower than the \twco\ line, especially at and close
to the \twco\ peak region along the bar. Moving away from this peak, the 
\twco\ linewidth drops and approaches that of \thco\ (see the dispersion maps
for \twco\ and \thco\ (Fig.\,\ref{disp}) and the spectra displayed in 
Fig.\,6). Especially from Fig.\,\ref{disp}c), which displays the ratio
of the \twco\ and the \thco\ velocity dispersion, it is clear that the 
velocity dispersion of \thco\ always is smaller than the velocity 
dispersion of \thco . The dispersions are similar (ratio $<2$) in the central
$\sim 10''$, where the \thco\ line is fairly wide ($\delta v_{\rm mom} \approx
60 - 80$\,\kms ) and in the southern \thco\ concentration, where the \twco\
transition becomes narrow ($\delta v_{\rm mom} \approx 10 - 30$\,\kms ). 
The ratio of the dispersion is high (3 -- 6) close to the \twco\ peak 
region along the bar,
where the \twco\ line is wide, while the \thco\ line is narrow.

The global \twco /\thco\ intensity ratio, i.e.\ the ratio we find
when comparing the total fluxes for the entire map, is quite high 
at $\sim 40$, certainly
much higher than what is typical for gas in galactic disks or even
`normal' galactic nuclei (see Table\,\ref{ratios} for typical values).
It is very unlikely that this ratio is an artifact of missing flux, since 
almost all flux in \twco\ is seen and we do not expect to miss more flux in
the the more compactly distributed \thco . However, lack of sensitivity
to \thco\ is a concern. 

To evaluate the possible magnitude of this effect,
we assume that \thco\ emission in a velocity range of 200\,\kms\ is present 
at the 2$\sigma$ level in all places where \twco\ is detected and we 
do not find \thco . This (unlikely) scenario would result in an additional 
flux of 10.8\,Jy\,\kms , i.e.\ roughly double our \thco\ flux and bring 
the global ratio down from 42 to 19, which we thus regard as a firm lower 
limit. We have also estimated the global value of \rco\ using the global
spectra of \twco\ and \thco , i.e.\ spectra constructed for the entire 
spatial extent of \twco\ emission. Here, \thco\ is only tentatively detected.
Using a 3$\sigma$ limit and a velocity range of 450\,\kms , we obtain a
ratio of \rco\,$\geq$\,30, in full agreement with the ratio determined 
using integrated fluxes.  

\begin{table*}
\caption{\label{ratios} $^{12}$CO/$^{13}$CO $1 \to 0$ line ratios (\rco )
for NGC\,7479 and typical galactic ratios. The
`cuts' refer to Fig.\,\ref{intcut}. Individual random errors are given 
in parentheses. The error in the scaling of all ratios due to calibration
uncertainties is estimated to be $\sim$ 20\%. }
\begin{tabular}{l|r|rr|rr}
  & & \multicolumn{4}{c}{Integrated Ratios} \\
Position  & Peak Ratio & Region of & Range$^{a)}$  & Region of  
& Range$^{a)}$ \\
  &   & $^{13}$CO line & [km\,s$^{-1}$] & $^{12}$CO line & 
[km\,s$^{-1}$]  \\
\hline 
global (bar and center) & -- & $\lapprox$\,42$^{b)}$ \\
\twco\ maximum intensity & 27(7) & 19(2) & 320 & 25(3) & 440  \\
\thco\ maximum intensity & 16(6) & 18(2) & 320 & 23(3) & 440  \\
cent.\ cond.\ (average)  & 22(8) & 22(4) & 400 & 23(4) & 440  \\
central region (OSO)     & 19(4) & 21(3) & 390 & 28(4) & 500  \\
cent.\ cond.\ (individual) & \multicolumn{4}{c}{10 -- 30$^{c)}$} \\
Peak of cut b & 27(9) & 25(5) & 220 & 29(7) & 400 \\
\twco\ peak of cut c & 19(5) & 17(4) & 60 & 40(15) & 340 \\
\thco\ peak of cut c (east) & 6.4(1.6) & 10(2) & 105 & 14(4) & 200 \\
\twco\ peak of cut d & 11(3) & 10(2) & 80 & 28(6) & 200 \\
\thco\ peak of cut d & 4.7(2.2) & 5.1(2.4) & 80 & 17(3) & 180 \\
\twco\ peak of cut e & 24(10) & 22(5) & 100 & 73(37) & 260 \\
\thco\ peak of cut e & 3.5(1.4) & 3.5(0.6) & 100 & 3.5(0.6) & 100 \\
northern \thco\ peak & 12(4) & 10(2) & 140 & 14(3) & 220 \\
\multicolumn{6}{c}{no \thco\ detections:} \\
offset +10$''$ along bar & -- & -- & --  & $>$15$^{d)}$ & 220  \\
offset $-17''$ along bar & -- & -- & --  & $>$25$^{d)}$ & 280  \\
offset $-21''$ along bar & -- & -- & --  & $>$25$^{d)}$ & 160  \\
\hline
galactic disk & \multicolumn{2}{c}{ $\sim$ 6$^{e)}$} \\
centers of `normal' galaxies & \multicolumn{2}{c}{13\,$\pm 6^{f)}$} \\
\hline 
\end{tabular} \\
a) The velocity range gives the region where the $^{12}$CO or $^{13}$CO
line is detected above the noise. \\  
b) Under extreme assumptions, this can be brought down to 20 (see text) \\
c) Range encountered in the central $10''$, both in peak and integrated
ratios \\
d) 3$\sigma$ limit calculated over the range where $^{12}$CO is visible. \\
e) Polk et al.\ \cite{polk+}; f) Aalto et al.\ \cite{aalto+} \\
\end{table*} 

In the central condensation, the variation in \rco\ goes along with 
the morphological shift described above. \rco\ ranges from $\sim 10$
close to the \thco\ `ridge' to $\sim 30$ at distances of 5$''$ or more from
it. The average ratio over the central condensation is fully 
consistent with 
the ratio determined from the single dish spectra taken toward the 
central position (Fig.\,\ref{sd}). Since we expect the central 
condensation to dominate the single dish flux
in the central 33$''$, this result confirms our prior 
conclusion that there is no missing flux in the interferometry maps. 

The even more drastic variation of \rco\ in the bar is 
evident from both Table\,\ref{ratios} and Fig.\,\ref{intcut}. 
The latter presents another easy way of visualizing the
changing line ratios. It shows cuts in integrated intensities (over a
range of 450\,\kms , chosen to include all \twco\ emission) along the 
bar minor axis at a number of offsets. Here, we focus on the southern
\thco\ extension. As the \thco\ emission shifts eastward from \twco\
when we move south along the bar, the line ratios change from values
around 25 in the central condensation to $\sim 40$ 
(integrated intensity) at the \twco\  maximum along the bar. The
ratios calculated using the peak flux and the velocity range of the 
\thco\ emission are often smaller than the \rco\ referring to the 
velocity range of the \twco\ line since in many places the \thco\ lines 
are narrower than the \twco\ lines. At the declination of the maximum
of the southern \thco\ distribution, \rco\ drops from 73$\pm$37 to 
3.5$\pm$0.6 (integrated) or 24$\pm$10 to 3.5$\pm$1.4 (peak) within $5''$! 
Interestingly, in this position, off the \twco\ peak, the widths of the
two transitions are identical. These values of \rco\ are put
into perspective by noting that  \rco\ of 6 -- 7 is typical for global 
ratios in galactic spiral arms (Polk et al.\ \cite{polk+}), which might
already include the contribution of some diffuse gas, while Galactic 
GMCs typically show \rco\ values of 3--5.  An \rco\ exceeding 30 is only seen
in a few luminous mergers (Casoli et al.\ \cite{cas+}, Aalto et al.\ 
\cite{aal2+}). 

The central gas surface density of NGC\,7479 
($\sim 2000 M_{\odot}$\,pc$^{-2}$, using
the `standard' CO to N(H$_2$) conversion factor for comparison purposes
only, see discussion below) is high enough to place the nuclear region in
the range of IR-bright starburst galaxies (Scoville \cite{scoville}),
even though ultraluminous IR galaxies (ULIRGs) can have core gas surface 
densities up to an order of magnitude higher (e.g.\ Bryant \& Scoville 
\cite{brysco}, who obtained exceeding 20000 $M_{\odot}$\,pc$^{-2}$ for 
NGC\,2623, Mrk\,231 and NGC\,6240, assuming a standard conversion factor). 
However, the gas surface density in the bar of NGC\,7479, where the most 
extreme values of \rco\ are reached, is far smaller than in luminous 
starbursts, pointing to a different dominant mechanism accounting for the 
high value of \rco .

The large changes of \rco\ have to indicate a very significant 
change in the properties of the emitting gas, scenarios for which will 
be discussed below

HCN\,$1 \to 0$ emission is detected only in the nucleus, with a 
\twco /HCN line ratio $\approx 20$.

\section{Discussion}

\subsection{Gas Masses Based on CO Isotopomers}

The total flux in our map {\em would} correspond
to an H$_2$ mass of $4 \cdot 10^{9}$\,\solmass\ {\em if} the `standard' CO to 
N(H$_2$) conversion factor (SCF = $X_{\rm CO} =2.3 \cdot 10^{20}$\,cm$^{-2}$
(K\,\kms)$^{-1}$, Strong et al.\ \cite{str+}) were applicable, 
which is, however, not the case (see discussion below). 
Instead of giving more masses based on the standard conversion 
factor, which are unlikely to be very meaningful, we compare
the `standard' mass derived from \twco \ to masses from \thco \ obtained
with the simple LTE assumptions of optically thin emission and a kinetic
gas temperature of 20\,K. {\em Globally\/}, i.e.\ avaraged over the entire 
map, the \thco -derived mass is a factor of 16 -- 27 (for \twco /\thco\  
abundance ratios between 30 and 50) lower than the SCF mass. If
the upper limit for the \thco \ total flux (see below) is taken, the
discrepancy decreases to 8 -- 13, still remaining very substantial. A 
mass discrepancy of $\sim10$ is also derived for the central condensation. 
In \twco\ peaks along the bar, the disagreement between the SCF and the 
\thco \ LTE mass reaches values exceeding 30, while in the \thco \ maximum 
in the southern bar, it comes down to $\sim 2$. Of course, these varying mass
discrepancies simply reflect the changing line intensity ratios discussed
above. 

The SCF mass still exceeds the \thco\ LTE mass by a factor 2 -- 10 if
the kinetic gas temperature is raised from 20\,K to 100\,K for the 
\thco\ - based estimate. Only in the \thco\ peak in the bar south of the 
central condensation the SCF mass drops to about half the \thco\ 
LTE mass under these conditions, which are, of course,
not realistic for a \thco\ condensation,
presumably consisting of dense gas. However, it seems that the SCF might be 
correct in this (and only in this) location, which has almost exactly 
the \twco /\thco\ line ratio found in Galactic GMCs. This can be considered
as a way to confirm the SCF for a normal spiral environment. 

Of course, assuming LTE conditions when deriving $N$(H$_2$) from \thco\  
could underestimate the true $N$(H$_2$) when non-LTE conditions apply.
Padoan et al.\ (\cite{pad+}) find a discrepancy of a factor  1.3 -- 7
between `LTE masses' and `true' masses 
in their cloud models. This discrepancy is smaller than the difference 
between 
the \thco\ LTE masses and the SCF masses we find both for the gas in the 
center and, even more pronounced, for the gas in \twco\ peaks along the bar 
away from center. Still, the \thco\ masses derived
for LTE should be considered as a lower limit to the real mass, while the 
SCF masses are upper limits. 

Simple non-LTE radiative transfer calculations (using the Large Velocity
Gradient (LVG) assumption) reproduce the results of Padoan et al.\ to within 
roughly a factor of two and can thus serve as a guide to possible scenarios.
These calculations show that the $^{12}$CO-to-H$_2$ conversion factor can be 
brought down significantly (by a factor of $\sim 6$) in low density gas 
($n({\rm H}_2) \gapprox  100$\,cm$^{-3}$). Then, high 
$^{12}$CO/$^{13}$CO 1 $\to$ 0 line intensity ratios of $20 - 30$ are  
predicted. Thus, this scenario points to the presence of `diffuse' gas,
discussed in detail in the next section.  

When the density is reduced even further, in the LVG models the 
$^{12}$CO-to-H$_2$ conversion factor rises again. The SCF can be recovered
for a H$_2$ density of $\sim 10$\,cm$^{-3}$, while \rco\ remains high.
We consider these extremely low densities required for the bulk of the gas
unlikely for the following reasons: To reach a column density that is 
sufficient to ensure shielding of the molecules against UV radiation and 
to be consistent with the observed brightness of the lines, even for an
extreme and very unlikely beam filling factor of unity, the models  
require an unrealistically low velocity dispersion for a given pathlength
($\leq 0.05$\,\kms\,pc$^{-1}$). This is in contradiction to the expectation 
that diffuse gas should be characterized by a large velocity dispersion (see
below). 
Another argument against extremely low density gas is the expected very low 
\twco (2$\to $1)/(1$\to $0) line intensity ratio ($< 0.3$), in contradiction
to single-dish (Sempere et al.\ \cite{sempere+}) and recent interferometric
observations of \twco (2$\to $1) (Baker \cite{baker}).

\subsection{Physical Conditions -- Diffuse Molecular Gas}

\begin{figure*}
\resizebox{18cm}{!}{\includegraphics{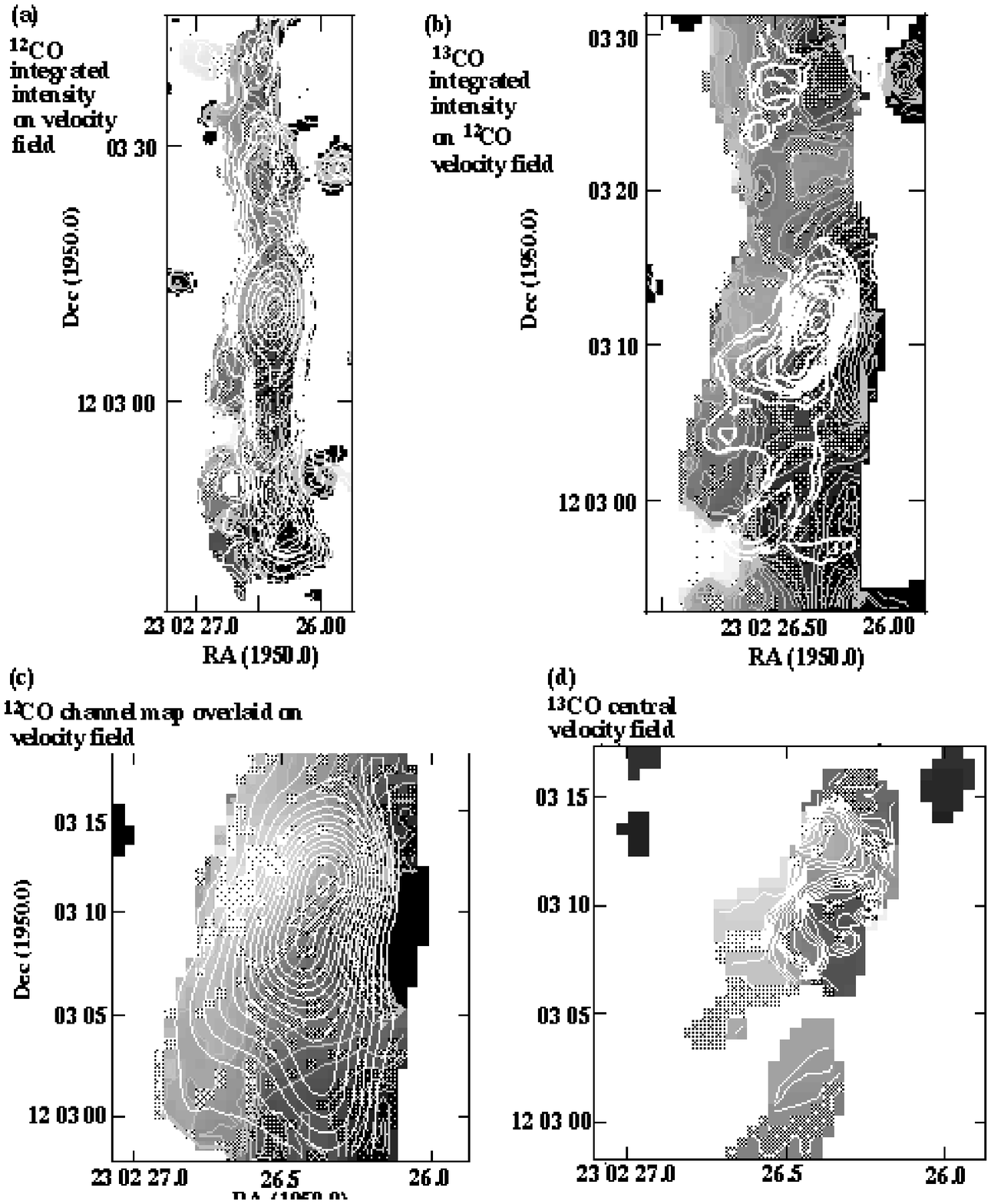}}
\caption{\label{vgrad} (a) The integrated \twco\ intensity (white contours)
superposed on the velocity field derived from\twco\ (grayscale and thin gray
contours). The velocity contours range from 2280\,\kms\ to 2440\,\kms\ 
and increase by 10\,\kms .
(b) The integrated \thco\ intensity (white contours) superposed 
on the velocity field from panel (a). The velocity contours
are the same as in panel (a) and thus facilitate a direct comparison of 
the two panels.
(c): \twco\ channel map for the velocity range 
2280\,\kms\ -- 2340\,\kms\ (white contours) overlaid on the \twco\ velocity 
field in the central part of the bar (grayscale and gray contours). The 
intensity contours are multiples of $2\sigma$ (5.9\,K\,\kms\ 
(1.24\,\jb\,\kms\ or 2$\sigma$). The velocity contours are the same as in
panel (a) . In grayscale, dark corresponds to higher velocities. 
(d): The central velocity field for \thco . The velocity contour
levels are the same as in panel (a). } 
\end{figure*}

In this section, we present simple scenarios that might explain the
variations in \rco . The key component of these scenarios is the
presence of a {\em diffuse or `intercloud' molecular medium (ICM)}.
This medium consists of low density ($n \lapprox 10^3$ cm$^{-3}$), 
gravitationally unbound, molecular gas. For a given column density, $N$, 
the \thco\ 1 $\to$ 0 line emission from gravitationally unbound gas will 
be weaker than that from self-gravitating clouds, because the velocity
dispersion, $\delta v$, is higher for unbound material and thus
$N/\delta v$ is lower for the unbound gas. A low $N/\delta v$ results 
in a low optical depth and, in low-density gas, also results in a low 
excitation temperature of the transition because of reduced radiative
trapping. Hence, the ICM will be 
difficult to detect in the \thco\ $1 \to 0$ line. One may see the
ICM as somewhat analogous to high latitude molecular clouds detected 
in the disk of our Galaxy. These are also low density structures which,
unlike most of the disk molecular clouds, are not in virial equilibrium.

While, under these conditions, \thco\ is
optically thin and subthermal, \twco\ reaches a moderate optical 
depth close to unity. Thus, \twco\ still radiates efficiently 
(and is self-shielding) while the \thco\ intensity falls off. This
moderate optical depth of the \twco\,$1 \to 0$ line is the main reason why
{\em the `standard' $X_{\rm CO}$ conversion factor cannot be 
applied to regions where a significant amount of diffuse gas is found\/}: 
the gas is neither optically thick nor virialized and the `standard' 
$X_{\rm CO}$ overestimates the molecular gas mass by up to an order 
of magnitude. This situation is encountered, e.g., in our Galactic
center (Dahmen et al.\ \cite{dah+}). 

\subsubsection{The Bar -- Outside the Central 10$''$ Diameter}

The large variations in \rco\ along the bar, with values ranging from
5 to $> 40$ (and even 73 in one extreme case) is best explained
by the presence of a large amount of diffuse (thin, warm, unbound)
ICM gas which dominates the \twco\ 1$\to$0 emission and is very 
distinctly different from the dense molecular clouds preferentially
detected by $^{13}$CO. In the southern part of the bar, where \thco\
is detected, the difference in linewidth is most dramatic close to the
\twco\ maxima. Here, the \twco\ lines have a width of $\sim 200$\,\kms ,
compared to $\sim 50$\,\kms\ for \thco . This explains the significant 
difference in \rco\ depending on whether integrated or peak intensities
are used (Table\,\ref{ratios}). The large linewidths of the \twco\ spectra
suggest large velocity dispersions at the \twco\ maxima along the bar. This
provides a strong argument in favour of a diffuse medium contrary to 
[$^{12}$CO]/[$^{13}$CO] abundance variations. While in a `standard' 
\twco\ emitting ISM, i.e.\ a medium where low $J$ \twco\ lines are 
optically thick, the [\twco ]/[\thco ] abundance ratio is not traced by
changes in \rco , a change in the \thco\ abundance itself could contribute
to changes in \rco . However, if such a variation reflects changes in
[$^{12}$C]/[$^{13}$C] with galactocentric radius, it should not exceed 
a factor of two (Wilson \& Matteucci \cite{wima}, Langer \& Penzias
\cite{lape}). Even a gradient this small and insufficient to explain
the observed changes in \rco\ is almost impossible 
to maintain in the bar due to efficient mixing (see e.g.\ Friedli et al.\
\cite{frie+} for model calculations, Martin \& Roy \cite{marroy} and 
Zaritsky et al.\ \cite{zar+} for empirical studies). Thus, we consider 
diffuse gas to provide the only viable 
explanation for the large variations in \rco\ we find along the bar of 
NGC\,7479.

Of course, \thco\ may be selectively photodissociated in the diffuse 
medium and thus really be underabundant, since it is not self-shielding. 
However, the efficiency of this mechanism also relies on the diffuse 
nature of the gas. 

According to the single dish map of Sempere et al.\ (\cite{sempere+}),
the \twco (2$\to$1)/(1$\to$0) ratio along the bar is $\sim 0.6$. This
is, of course, a global value which is likely to vary at higher
resolution. It is compatible with cold gas at moderate density 
($10^3 - 10^4$\,cm$^{-3}$) or warm, thin gas 
(density a few 100\,cm$^{-3}$). Given the evidence discussed aboved, 
we obviously favour the latter explanation for this line ratio. 

The fact that the \thco\ intensity maximum is clearly offset along the
bar minor axis from the \twco\ maximum in the southern bar indicates
that \thco\ traces a largely different component. The relation of
the two tracers to dust, shocks and star formation will be discussed
further in Section\,4.3.

\subsubsection{The Inner Region -- Inside the Central 10$''$ }

At first glance, the situation in the center of NGC\,7479 might be taken
to argue against the ICM scenario outlined above, since the HCN is 
coincident with the \twco\ maximum and offset from \thco . A situation like
this was sometimes seen as suggesting abundance changes for the CO
isotopomers (e.g.\ Casoli et al.\ \cite{cas+}), since HCN requires high 
gas densities of at least $10^4$\,cm$^{-3}$ to be significantly 
excited and may be identified with the region where most of the mass 
resides. In a turbulent environment like the nuclear region of NGC\,7479,
we consider the alternative {\em explanation of a kinetic temperature
gradient between the \thco\ and HCN peaks} to be  more likely. For the high
gas densities we expect in the central region (i.e.\ $n$(H$_2) > 
10^4$\,cm$^{-3}$), the higher kinetic temperature 
would occur at the HCN peak rather than the \thco\ peak because the 
brightness of the \thco\ 1$\to$0 line is a decreasing function of kinetic 
temperature (for temperatures above 8\,K in the LTE, optically thin limit). 

The HCN molecule has a high dipole moment and is thus much more 
sensitive to density variations than \twco\ and \thco . Thus the HCN peak 
is not expected to coincide with the \thco\ peak if the latter is at lower
densities than the HCN\,1$\to$0 critical density. This is true even if the 
HCN\,1$\to$0 transition is optically thin, due to the spreading of the 
molecules over many rotational levels that occurs at high densities (i.e.\
$10^6$\,cm$^{-3}$ for HCN) because of the high  collision rates at these 
densities. 

We can give an order-of-magnitude estimate of the gas density that is
likely to occur at the HCN peak: Having the bulk of the gas in the central 
beam at densities of $\sim 10^6$\,cm$^{-3}$ requires a very low 
volume filling factor of $\lapprox 10^{-4}$, which seems unlikely 
in the center of a galaxy undergoing a mild starburst. Also, very high 
densities suggest a velocity dispersion per pathlength that is too high 
to be realistic (several 100\,\kms\,pc$^{-1}$). If, on the other hand, 
the density was low enough for HCN to be subthermally excited, the 
HCN\,1$\to$0 line would be optically thick, because the molecules would 
pile up in the $J=0$ level. However, in this case (at peak densities of 
$\sim 10^4$\,cm$^{-3}$) we would expect the emission to be more widespread. 
Low densities close to $10^3$\,cm$^{-3}$ lead to very subthermal 
conditions and very weak HCN\,1$\to$0 emission. Thus, we consider 
$\sim 10^5$\,cm$^{-3}$ to be a reasonable density at the HCN peak.  

In summary, since the HCN distribution is 
confined closely to the \twco\ peak, we expect this region to also be 
a peak in density (but not necessarily in column density), in addition to 
being a peak in \TKIN . In contrast, the \thco\,1$\to$0 emission will 
preferentially trace cool and/or lower density gas.

A kinetic temperature gradient was also used to
explain the different spatial extents of the \thco (3$\to$2) and HCN
emission in NGC\,253 (Wall et al.\ \cite{ww91}). The kinetic temperature
gradient in NGC\,7479 leads to the prediction that the \thco 
(2$\to$1)/(1$\to$0) ratio should be higher in the HCN peak than in the
\thco\ peak, as is seen in the nucleus 
of IC\,694 in the Arp\,299 merger (Aalto et al.\ \cite{aalarp}). It also
means that the nuclear \twco(2$\to$1)/\twco(1$\to$0) ratio is expected to be 
high, at least close to unity; this seems to be the case (Baker 
\cite{baker}). \thco\ 2$\to$1 observations will be crucial to further
investigate the gas properties in the central 1.5\,kpc of NGC\,7479
and make comparisons to luminous mergers, where Casoli et al.\ 
(\cite{cas+}) and Taniguchi et al.\ (\cite{tani+}), based on data by
Aalto et al.\ \cite{aalto+} and Casoli et al., have argued that
\thco\ 2$\to$1 is also relatively faint and thus \thco\ may be 
depressed, at least in some cases. 

We note, however,
\thco\ 1$\to$0 is not as faint at the \twco\ maximum as under the extreme
conditions of mergers: Fig.\,\ref{spcen} shows the line to be as strong 
as HCN -- it just peaks in a different position. The central \twco /HCN
line ratio of $\sim 20$ is on the high side if compared to the range 
found in the central region of a number of galaxies, where 10 is more 
typical; however the scatter is very large even among barred starburst
spirals (Contini et al.\ \cite{con+}). Still, a value of 20 may be
an indication that the amount of warm dense gas in the nuclear region
of NGC\,7479 is only moderate and confined to a small region.    

Diffuse gas is likely to be also important even in the central condensation
of NGC\,7479. This is indicated by the high values \rco\ reaches away
from the curved `\thco - ridge' in the central condensation. In these
areas, the \thco\ lines are much narrower than the \twco\ transitions
(150\,\kms\ -- 300\,\kms\ for \twco\ versus 50\,\kms\ -- 80\,\kms\ for
\thco). These differences in line shape are also obvious from the
global single dish spectra (Fig.\,\ref{sd}). 

Diffuse molecular gas has been detected in the centers of a large 
number of barred and starburst galaxies such as IC\,342 
(Downes et al.\ \cite{dow+}) or NGC\,1808 (Aalto et al.\ \cite{aal94}) and,
of course, in the bar region of the Milky Way (Dahmen et al.\ \cite{dah+}). 
It has also been detected in the center of the elliptical galaxy 
NGC\,759 (Wiklind et al.\ \cite{wik+}) where a low line 
ratio of 0.4 is indicative of a two component medium made up of dense, 
cold clumps embedded in a warm, diffuse molecular medium. A scenario 
where \twco\ and \thco\ emission arises from separate components was
suggested for IC\,342 as early as 1990 by Wall
\& Jaffe. 
The idea that diffuse gas characterized by \twco\ emission of
moderate optical depth is important even in regions of
extreme star formation, usually associated with large amounts of dense 
molecular gas, has recently gained credibility even for ultraluminous
infrared galaxies (Aalto et al.\ \cite {aalto+}, 
Downes \& Solomon \cite{dowsol}). 

It is interesting to note that the \thco\ ridge in the central
condensation is precisely aligned with the steepest velocity gradient
(derived from \twco , Fig.\,\ref{vgrad}).

\subsection{Production of an ICM by tidal disruption and cloud collisions}

\begin{figure}
\rotatebox{90}{
\resizebox{6cm}{!}{\includegraphics{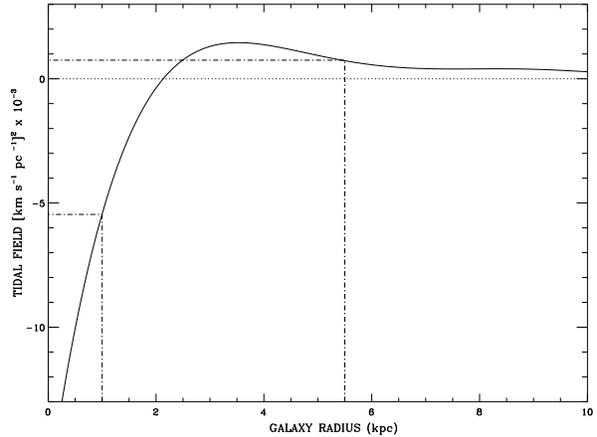}}}
\caption{\label{tidal} Tidal field per unit mass and across unit length
(see text for explanation) along the bar. The dotted lines indicate
the tidal field at radii 1\,kpc and 5.5\,kpc, which are the semiminor
and semimajor axes of the bar. This illustrates the change in tidal force 
that a cloud on an elongated bar orbit can experience.} 
\end{figure}

Cloud disruption due to the tidal field along a bar is one mechanism 
that could produce a diffuse ICM, especially close to the central 
condensation. Another possible source of the ICM are off-center 
cloud collisions. 

(i) {\em Cloud evaporation due to the tidal field \/}: 
As a bound cloud moves in a bar potential or elliptical orbit, the tidal
field across it will vary in time. This produces both internal heating of 
the cloud and clump evaporation from the outer regions of the cloud 
(Das \& Jog \cite{dasjog}). The evaporated cloud mass will become part of 
the low density, molecular ICM, raising \rco . 

To get a lower estimate of how the tidal field changes over the bar, we 
first determined the potential of the galaxy from the deprojected K image 
data (Combes, private communication). The potential was azimuthally
averaged to determine $\Phi(r)$, which was fitted with a polynomial
function that was used to derive a rotation curve for the galaxy. The
mass to luminosity ratio $(M/L)$ was chosen so as to obtain the best fit of
the derived rotation curve to that observed. We then used this
azimuthally averaged  potential to calculate the tidal field over the
galaxy. Though this does not give the exact variation of the tidal field 
across a cloud in the bar potential, it will give a first appoximation as to 
how the tidal field varies. In Fig\,\ref{tidal} we have plotted the 
tidal field per unit mass, across unit length, 
$t_f$ = $(\partial F/\partial r)\ ({\rm km}\,{\rm s}^{-1}\,{\rm pc}^{-1})^2$ 
against radial distance $r$ in kpc.  

With a projected bar length of approximately 10.4\,kpc and width 
1.6\,kpc (Fig.\,\ref{12co}), an inclination angle of 41$^{\circ}$ 
and a position angle 25$^{\circ}$ for the bar (Sempere at al.\
\cite{sempere+}), the deprojected bar length is about 11\,kpc and its 
width is about 2\,kpc (Arnaboldi et al.\ \cite{arna+}). 
So at the semimajor axis distance of r=5.5 kpc, 
$t_f = 0.8 \cdot 10^{-3}\ ({\rm km}\,{\rm s}^{-1}\,{\rm pc}^{-1})^2$
and at the semiminor axis 
distance of $r=1$\,kpc,  $t_f = -5.5 \cdot 10^{-3}\
({\rm km}\,{\rm s}^{-1}\,{\rm pc}^{-1})^2 $. 
Thus, the tidal field changes from being disruptive at the bar ends 
to compressive close to the bar center. This considerable
change in tidal field, which is steepest in the central region, may 
produce significant cloud evaporation, thus leading to large amounts
of diffuse gas.  

\begin{figure}
\resizebox{8cm}{!}{\includegraphics{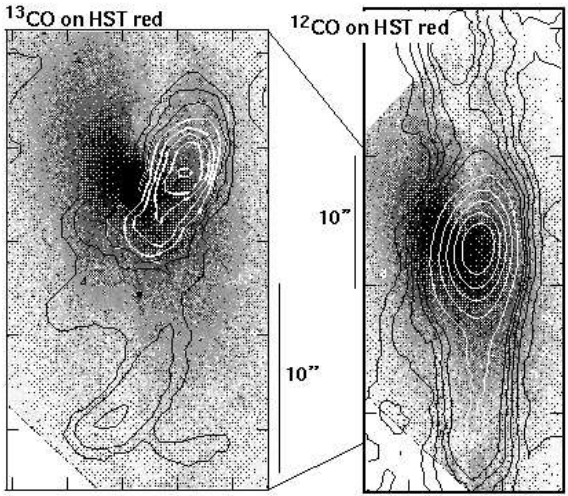}}
\caption{\label{hst} Superposition of the \twco\ distribution (right)
and the \thco\ distribution on an archival HST (Planetary Camera) red 
image that shows the shape of the dust lanes.} 
\end{figure}

(ii) {\em Cloud Collisions\/}: Hydrodynamic simulations have shown that 
off-center cloud collisions lead to gas being sheared off the colliding 
clouds, forming trailing extensions in the interstellar medium; the 
gas then dissipates into the ISM (Hausman \cite{haus}). This gas may 
form part of the diffuse ICM.
In bars, the crowding of closed $x_1$ orbits at the bar ends enhances the
cloud collision frequency signigicantly and also produces shocked gas.
Collisions also cause the clouds to lose angular momentum and sink inwards
leading to the central buildup of gas. This increased concentration of gas
in the centers of starburst galaxies and barred galaxies means the rate of
formation of diffuse gas in these galaxies will also be high.
NGC\,7479 has a strong bar 
and may have undergone a recent minor merger (Laine \& Heller 
\cite{laihel}). Thus, the cloud collision rate should be high and there 
should be a considerable amount of diffuse gas in the central 
$\sim 1.5$\,kpc.

Distinguishing between the two mechanisms observationally is not easy.
The diffuse gas produced by cloud collisions may often be accompanied
by shocked, dense gas. Tidally evaporated gas will not have this association.
However, diffuse gas is expected to spread over the bar more quickly than
dense gas, which will make it difficult to determine its origin. Still,
in the future sensitive, high resolution observations of a shock tracer 
like SiO may be useful to decide this question.  

\subsection{The Relation between Tracers of Shocks and Star Formation}

The S-shaped, complex velocity field of the gas in the bar of NGC\,7479
(see Fig.\,\ref{vgrad}) has been convincingly explained by e.g.\ Laine et 
al.\ (\cite{laine+}) as the result of gas streaming motions in a strong bar
potential. They further argue that H$\alpha$, dust lanes, a large 
velocity gradient and strong \twco\ emission more or less coincide.
These are all taken to trace, in some sense, shocks, gas compression and 
star formation.

Including \thco\ and the varying line ratios, indicative of at least two 
different components of molecular gas, (and seeing all the \twco ) leads to
a more complex picture. 
In Fig.\,\ref{hst}, we overlay the \twco\ (right) and \thco\ (left)
intensity distributions on an HST red image retrieved from the archive. 
In the north, where no \thco\ is detected outside the central region, \twco\
follows the strong dust lane almost perfectly. In the south, the visual
dust lane is weaker or more diffuse and the correlation with the 
\twco\ distribution is much less evident. However, the southern 
\thco\ peak coincides with a (faint) peak in a dust lane. South of the
center, \rco\ grows when we move upstream (west) across the bar 
(assuming that the spiral arms are trailing). In other words,
the \thco\ concentration is located downstream from the \twco\ maximum, and
possibly also downstream from the peak of the shock, if we identify the
\twco\ maximum with the shockfront. This may be justified by the close
coincidence of the \twco\ maximum with the optical dust lane in the 
north and also by the excellent agreement Laine et al.\ (\cite{laine+})
find between \twco\ peaks (both in the north and in the south) and the 
maxima in a $J-K$ NIR color map that is an even better indicator of dust 
than an optical image.  

In the central concentration, the course of the dust lanes is not clear.
However, here it is the \thco\ that very clearly follows the steepest velocity
gradient (Fig.\,\ref{vgrad}b), while \twco\ is distributed fairly 
independently of the velocity gradient. \twco -emitting gas at a velocity
between 2280\,\kms\ and 2340\,\kms\ traces the velocity gradient somewhat 
more closely than gas at other velocities (Fig.\,\ref{vgrad}c), but even 
in this restricted velocity range the \twco\ emission does not follow the 
velocity field as clearly as the \thco\ emission does. A number of spectra
from the central 15$''$ have their emission peak close to 2300\,\kms , and
a narrow emission component may be associated with this peak (faintly visible
in Fig.\,4).

The velocity field based on \thco\ is displayed in Fig.\,\ref{vgrad}d). 
It is even more dramatically
S-shaped than the \twco\ velocity field. We further note that the \twco\
velocity field determined by Laine et al.\ closely resembles our \thco\ 
velocity field, and that their \twco\ distribution in the central 15$''$
appears more bent than ours.

This evidence leads us to the following scenario: The \thco\ emission and
a part of the \twco\ emission, especially at mid-velocities, traces dense
or at least high column density gas that follows the velocity field and is
compressed in the region with the steepest velocity gradient. We suspect that 
this region coincides with the central dust lane, which is, however, not
clearly visible in the HST image. In terms of \twco\ emissivity, this gas
component is not dominant enough to clearly reduce \rco . Also,
the dust lane may be fairly narrow and unresolved in our data. Thus, this
component is seen most clearly in the \thco\ data, but Laine et al.\ also
trace it in \twco\ since they are not sensitive to much of the extended 
emsission. The \twco\ and \thco\ distributions in our maps deviate because
the \twco\ emission, especially but not exclusively at non-central velocities,
is due to diffuse gas that is far less confined to the shock region in the
central $10'' - 15''$ ($\sim 2\,kpc)$. The dramatic S-shape in the \thco\
velocity field (Fig.\,\ref{vgrad}d) indicates that the strong inner
bar of NGC\,7479 is the feature dominating the kinematics of the 
denser gas in this central region. As discussed earlier (Fig.\,2), a rotating
disk or gas on $x_2$-orbits, perpendicular to the main bar axis, is 
interacting with the bar on the scale of a few hundred pc at the very center.

It is conceivable that the diffuse gas rises to higher altitudes above the
plane of the galaxy than the dense gas, which may be less turbulent and 
more strongly dominated by the bar potential, which is defined by the stellar
potential, since most of the dynamical mass is almost certain to be in the 
form of stars. 
    
Outside the central 2\,kpc, the relation between the various tracers changes:
both the detached northern and the southern \thco\ peaks are found close to, 
but not coincident with, regions of steep velocity gradients 
(Fig.\,\ref{vgrad}). The dust lane (and presumably the region of the strongest
shock) is now traced well by the \twco\ emission. Possibly, the larger overall
amount of gas in the central 2\,kpc allows a diffuse component to spread 
throughout this region, while in the outer bar the gas associated with the
shock remains diffuse.  

It seems possible that the \thco\ peaks in the outer bar indicate 
conditions that are more quiescent and favourable to the formation of
bound clouds and ultimately star formation activity. The low value of \rco\
in these complexes suggests that their properties are similar to GMCs in
the spiral arms of normal galaxies. The cuts presented by Sempere et al.\
(\cite{sempere+}) 
indicate that the H$\alpha$ emission peaks along with the \thco\ in the 
southern condensation, i.e.\ is offset from the \twco\ and main dust lane
in this area. If the H$\alpha$ traces star formation, this shift away from
a region that seems to be totally dominated by diffuse gas may reflect 
that star formation occurs wherever the gas density is high enough, while
the proximity to the shock front may play a secondary role. The fact
that the \thco\ condensations are close to (offset $1'' - 2''$) strong 
(\twco ) velocity gradients need not be a contradiction to quiescent 
conditions. The peaks of the condensations are downstream of the strongest
gradient by several hundred pc. In addition, the interiors of the 
condensations could be quiescent while the condensations themselves are 
following a strong overall large-scale flow. 
This question,
however, requires further study on smaller spatial scales.  

Offsets between molecular gas concentrations and dust lanes 
have recently been reported by Rand et al.\ (\cite{rand+}) for a part of 
a spiral arm in M\,83. They used \twco\ as a tracer, but see only 
2\% -- 5\% of the single dish flux in their interferometer map.
Thus, they probably filter out the diffuse component, so their \twco\
data might trace a dense, clumpy component similar to what we
see in \thco . 

\section{Conclusions}

We have mapped the barred spiral galaxy interferometrically in the
$J = 1 \to 0$ transitions of \twco , \thco\ and HCN. Our main results are
as follows:

\begin{enumerate}
\item The \twco\ map shows a continuous gas distribution all allong the bar.
Comparison to single dish observation shows that the interferometer does
not miss a significant amount of flux. The velocity field derived from 
the $^{12}$CO map is complex, showing the S-shaped isovelocity contours 
typical of noncircular gas orbits in a strong bar.

\item A high velocity feature is identified close to the center. This may 
be a ring associated with an Inner Lindblad resonance, a tilted rotating
disk fed directly by mass infall along the bar or even an inner bar decoupled
from the main bar. 

\item \thco\ emission is detected in the central condensation, the southern 
part of the bar and a single location in the northern part of the bar.
HCN emission is only detected from the center. 
The \thco\ central emission is offset by $\sim 2''$ from the $^{12}$CO
and HCN intensity maxima, which are coincident. Along the bar, the most 
prominent peak of southern \thco\ condensation is also clearly offset 
from the \twco\ distribution.

\item The \twco /\thco\ line intensity ratio, \rco , varies dramatically. 
Globally (9\,kpc $\times$ 2.5\,kpc), \rco\ is 20 -- 40, 
a high value compared to typical
ratios found in the disk component of galaxies or even central regions of
normal galaxies. This indicates a prominent contribution of diffuse, unbound
molecular gas with a moderate optical depth in the \twco (1$\to$0) transition
in both the bar and the center of NGC\,7479.

\item On smaller sizescales of $\sim 750$\,pc, 
\rco\ exceeds 30 in large parts of the bar, reaching
values usually found in starburst mergers. Since values as low as 5 are
also found in the bar, close to the \thco\ condensation,  and since a
bar environment is very well mixed, 
we discard an underabundance of the
$^{13}$C isotope as a possible explanation of the very high \rco\ found
in many places. Instead, this is explained by a dominant component of 
diffuse gas, readily produced by either tidal disruption or cloud
collisions in the bar potential. The large variation in \rco\ is reflected
by large changes of likely values of the conversion factor from \twco\
intensity ro H$_2$ column density. In the central 1.5\,kpc, the Galactic
`standard' conversion factor (SCF) overestimates the gas mass by a factor of
up to 10; in \twco\ peaks along the bar the discrepancy is even larger.
Only in a \thco\ complex in the bar we find the SCF to be correct.  

\item The offset in the central HCN (and \twco ) peak from the \thco\ peak
can be attributed to a gradient in kinetic temperature in which the highest 
gas kinetic temperature is at the position of the HCN peak. This leads to 
the prediction that the \thco (2$\to$1)/(1$\to$0) intensity ratio should be
higher at the HCN peak than at the \thco\ peak. 

\item The region along the bar where \rco\ is small might be an area where
the conditions are more quiescent, which is also indicated by the narrowness
of the lines, both in \twco\ and \thco , found here. If the \twco\ ridge
along the bar, which coincides closely with the dust lanes, is taken to be
the location of the bar shock, the \thco\ condensation is behind or downstream
this shock, possibly in a region where the disrupted (but also compressed)
gas emerging from the shock can form bound molecular complexes. However, 
the region where the velocity gradient is steepest along the bar does not
coincide exactly with either the \twco\ or the \thco\ distribution.
In the center, \thco\ traces the steepest velocity gradient much 
more closely than $^{12}$CO. Thus, the relation between the molecular
tracers and the shock is complex.

\end{enumerate}

\acknowledgements{The OVRO mm-array is supported in part by NSF grant
AST 9314079 and the K.T.\ and E.L.\ Norris Foundation.}

\end{document}